\documentclass[12pt,a4paper]{article}

\newcommand{\st}{\mathrm{STr}}
\newcommand{\td}{\tilde{D}}
\newcommand{\tq}{\tilde{Q}^{-1}}
\newcommand{\la}{\lambda}
\newcommand{\pa}{\partial}
\newcommand{\tr}{\mathrm{Tr}}
\newcommand{\bd}{\bar{D}}

\begin{document}

\begin{flushright}
{ }
\end{flushright}
\vspace{1.8cm}

\begin{center}
 
\textbf{\Large Janus Field Theories from Non-Linear \\
BF Theories for Multiple M2-Branes}
\end{center}
\vspace{1.6cm}
\begin{center}
 Shijong Ryang
\end{center}

\begin{center}
\textit{Department of Physics \\ Kyoto Prefectural University of Medicine
\\ Taishogun, Kyoto 603-8334 Japan}
\par
\texttt{ryang@koto.kpu-m.ac.jp}
\end{center}
\vspace{2.8cm}
\begin{abstract}
We integrate the nonpropagating $B_{\mu}$ gauge field for the non-linear
BF Lagrangian describing $N$ M2-branes which includes terms with even
number of the totally antisymmetric tensor $M^{IJK}$ in arXiv:0808.2473 
and for the two-types of non-linear BF Lagrangians which include 
terms with odd number of $M^{IJK}$ as well in arXiv:0809:0985. 
For the former Lagrangian we derive directly the DBI-type 
Lagrangian expressed by the $SU(N)$ dynamical $A_{\mu}$ gauge 
field with a spacetime dependent coupling constant, while for the 
low-energy expansions of the latter Lagrangians the $B_{\mu}$ 
integration is iteratively performed. The derived Janus field
theory Lagrangians are compared.

\end{abstract}
\vspace{3cm}
\begin{flushleft}
March, 2009 
\end{flushleft}

\newpage
\section{Introduction}

Inspired by Bagger and Lambert \cite{BL} and Gustavsson \cite{AG} (BLG) 
who constructed the worldvolume theory of multiple coincident 
M2-branes following earlier works \cite{JS,BH}, the multiple 
M2-branes have been extensively studied. The BLG theory is described by a
three-dimensional $\mathcal{N}=8$ superconformal Chern-Simons gauge theory
with manifest SO(8) R-symmetry based on 3-algebra with a positive
definite metric, that is, the unique nontrivial $\mathcal{A}_4$ 
algebra \cite{GP}. However, this Chern-Simons gauge theory expresses two
M2-branes on a $R^8/Z_2$ orbifold \cite{MR}. 

A class of models based on 3-algebra with a Lorentzian
metric have been constructed by three groups \cite{GMR,BGTV,HIM}
where the low-energy worldvolume Lagrangian
of $N$ M2-branes in flat spacetime is described by a three-dimensional
superconformal BF theory for the $su(N)$ Lie algebra.
Using a novel Higgs mechanism of ref. \cite{MP} the BF membrane theory
has been shown to reduce to the three-dimensional maximally supersymmetric
Yang-Mills theory whose gauge coupling is the vev of one of the scalar
fields \cite{GMR,HIM,HIS}. For the prescription of the ghost-like scalar
fields a ghost-free formulation has been proposed by introducing a new 
gauge field for gauging a shift symmetry and then making the gauge 
choice for decoupling the ghost state \cite{BLS,GGRV,HV}.
In ref. \cite{EMP} starting from the maximally supersymmetric 
three-dimensional Yang-Mills theory and using a non-Abelian duality
transformation due to de Wit, Nicolai and Samtleben (dNS) \cite{NS},
the Lorentzian BLG theory has been reproduced.

The relation between the $\mathcal{N}=6$ superconformal 
Chern-Simons-matter theory \cite{ABJM} and the $\mathcal{N}=8$ 
Lorentzian BLG theory has been studied \cite{HISZ,JBL,PW,AT}.
The various investigations related with the BLG theory have been
performed \cite{HHM,AM,HL,HMS}

There has been a construction of a manifestly $SO(8)$ invariant
non-linear BF Lagrangian for describing the non-Abelian dynamics of the
bosonic degrees of freedom of $N$ coincident M2-branes in flat spacetime,
which reduces to the bosonic part of the BF membrane theory for $SU(N)$
group at low energies \cite{IR}. This non-linear Lagrangian is an 
extension of the non-Abelian DBI Lagrangian \cite{AAT,RM} of $N$
coincident D2-branes and includes only terms with even number of the
totally antisymmetric tensor $M^{IJK}$. Further, two types of 
non-linear BF Lagrangians have been presented such that they include
terms with even and odd number of $M^{IJK}$ \cite{MG}.
A different kind of non-linear gauged M2-brane Lagrangian has been
proposed for the Abelian case \cite{JK}.

As a related work, it has been shown that starting with the 
$\mathcal{N}=8$ supersymmetric Yang-Mills theory on D2-branes and 
incorporating higher-derivative corrections to lowest nontrivial order,
the Lorentzian BF membrane theory including a set of derivative 
corrections is constructed through a dNS duality \cite{MAM}
(see \cite{LLX}). The higher-derivative corrections to the Euclidean
$\mathcal{A}_4$ BLG theory have been determined \cite{BEMP} 
by means of the novel Higgs mechanism 
and also shown to match the result
of \cite{MAM}. The couplings of the worldvolume of  multiple 
M2-branes to the antisymmetric background fluxes have been investigated by
using the low-energy Lagrangian for multiple M2-branes \cite{LW,MGA}
as well as the non-linear BF Lagrangian \cite{MAG}. There have been
proposals for the non-linear Lagrangians for describing the 
M2-brane-anti-M2-brane system \cite{MRG} and the unstable
M3-brane \cite{JJK}.

We will perform the integration over the redundant 
$B_{\mu}$ gauge field for the non-linear BF Lagrangians of ref. \cite{IR}
and ref. \cite{MG}, to see how the Lagrangians are described by the
dynamical $A_{\mu}$ gauge field. We will carry out the $B_{\mu}$ 
integration directly for the non-linear Lagrangian of 
ref. \cite{IR}, while the $B_{\mu}$ integration will be iteratively
performed for the two types of non-linear 
BF Lagrangians of ref. \cite{MG}.
These three $B_{\mu}$ integrated Lagrangians will be compared.

\section{Non-linear BF Lagrangian with even number \\of $M^{IJK}$}

We consider the non-linear BF Lagrangian for $SU(N)$ group which 
describes the non-Abelian dynamics of the bosonic degrees of freedom of
$N$ M2-branes in flat spacetime \cite{IR}
\begin{eqnarray}
L&=& - T_2 \st\left( \sqrt{- \det\left(\eta_{\mu\nu} + 
\frac{1}{T_2}\td_{\mu}X^I\tq_{IJ}\td_{\nu}X^J \right) }
(\det \tilde{Q})^{1/4} \right) \nonumber \\
 &+& \tr\left( \frac{1}{2}\epsilon^{\mu\nu\la}B_{\mu}F_{\nu\la}
\right) + (\pa_{\mu}X_-^I - \tr(X^IB_{\mu}))\pa^{\mu}X_+^I \nonumber \\
&-& \tr \left( \frac{X_+\cdot X}{X_+^2} \hat{D}_{\mu}X^I\pa^{\mu}X_+^I
- \frac{1}{2} \left(  \frac{X_+\cdot X}{X_+^2} \right)^2
\pa_{\mu}X_+^I\pa^{\mu}X_+^I \right),
\label{la}\end{eqnarray}
where $X_+^2 = X_+^IX_+^I$ and the M2-brane tension $T_2$ is related to 
the eleven-dimensional Planck length scale $l_p$ as $T_2 = 
1/(2\pi)^2l_p^3$. The two non-dynamical gauge fields $A_{\mu}, B_{\mu}$
and the scalar fields $X^I \;(I=1,\cdots,8)$ are in the adjoint 
representation of $SU(N)$ and $X_{\pm}^I$ are $SU(N)$ singlets.
The covariant derivative $\td_{\mu}$ is defined by
\begin{equation}
\td_{\mu}X^I = \hat{D}_{\mu}X^I - \frac{X_+\cdot X}{X_+^2}\pa_{\mu}
X_+^I, \;\hat{D}_{\mu}X^I = D_{\mu}X^I - X_+^I B_{\mu}, \;
D_{\mu}X^I = \pa_{\mu}X^I + i[A_{\mu},X^I]
\end{equation}
and the SO(8) tensor $\tilde{Q}^{IJ}$ is given by 
\begin{equation}
\tilde{Q}^{IJ} = S^{IJ} + \frac{X_+^I X_+^J}{X_+^2}(\det S - 1),
\hspace{1cm} S^{IJ} = \delta^{IJ} + \frac{i}{\sqrt{T_2}}
\frac{m^{IJ}}{\sqrt{X_+^2}},
\label{qs}\end{equation}
where $m^{IJ}$ is expressed as
\begin{equation}
m^{IJ} = X_+^K M^{IJK}, \hspace{1cm} M^{IJK} = X_+^I[X^J,X^K] + 
X_+^J[X^K,X^I] + X_+^K[X^I,X^J].
\end{equation}
In (\ref{la}) $\tq_{IJ}$ denotes the matrix inverse of $\tilde{Q}^{IJ}$
and STr is the symmetrized trace \cite{AAT}. The non-linear Lagrangian 
$L$ is invariant under the obvious global SO(8) transformation and the
$SU(N)$ gauge transformation associated with the $A_{\mu}$ gauge 
field, and further the non-compact gauge transformation associated with
the $B_{\mu}$ gauge field
\begin{equation}
\delta X^I = X_+^I\Lambda, \hspace{1cm} \delta B_{\mu} = D_{\mu}\Lambda,
\hspace{1cm} \delta X_+^I = 0,\hspace{1cm} \delta X_-^I = \tr(X^I\Lambda).
\end{equation}
The terms except for the first non-linear term and the second BF-coupling
term in (\ref{la}) are added to have consistency with the low-energy 
Lagrangian. In the non-linear Lagrangian $L$ only the symmetric part of
$\tq_{IJ}$ is taken into consideration.

We introduce a Lagrange multiplier $p$ to rewrite the square root term 
in (\ref{la}) as
\begin{eqnarray}
&-& T_2 \sqrt{-\det \left(\eta_{\mu\nu} + \frac{1}{T_2} \td_{\mu}X^I
\tq_{IJ}\td_{\nu}X^J \right) }(\det \tilde{Q})^{1/4}
\nonumber \\
&\rightarrow&  \left( 
\frac{T_2^2}{2p} \det(\eta_{\mu\nu} + \frac{1}{T_2} \td_{\mu}X^I
\tq_{IJ}\td_{\nu}X^J ) - \frac{p}{2} \right)(\det \tilde{Q})^{1/4},
\label{sp}\end{eqnarray}
where a matrix can be treated as a c-number within the 
symmetrized trace. Owing to $\tq_{IJ} = \tq_{JI}$
the relevant tensor is rearranged as
\begin{equation}
\eta_{\mu\nu} +  \frac{1}{T_2} \td_{\mu}X^I\tq_{IJ}\td_{\nu}X^J =
g_{\mu\nu} + \tilde{B}_{\mu}\tilde{B}_{\nu},
\label{gb}\end{equation}
where
\begin{eqnarray}
g_{\mu\nu} &=& \eta_{\mu\nu} + \frac{1}{T_2} \bd_{\mu}X^I
\tilde{P}_{IJ}\bd_{\nu}X^J, \hspace{1cm} \bd_{\mu}X^I = D_{\mu}X^I -
\frac{X_+\cdot X}{X_+^2}\pa_{\mu} X_+^I, \nonumber \\
\tilde{B}_{\mu} &=& \sqrt{\frac{X_+^I\tq_{IJ}X_+^J}{T_2} } \left(
B_{\mu} - \frac{\bd_{\mu}X^I\tq_{IJ}X_+^J}{X_+^K\tq_{KL}X_+^L} \right)
\label{tb}\end{eqnarray}
with \begin{equation}
\tilde{P}_{IJ} = \tq_{IJ} - \frac{\tq_{IK}X_+^KX_+^L\tq_{LJ}}
{X_+^M\tq_{MN}X_+^N},
\label{tp}\end{equation}
which is orthogonal to $X_+^I$ as $X_+^I\tilde{P}_{IJ}=0$.

The expression (\ref{sp}) together with (\ref{gb}) is quadratic in 
$B_{\mu}$ so that the equation of motion for the auxiliary field
$B_{\mu}$ is given by
\begin{equation}
g^{\mu\nu}\left( B_{\nu} - \frac{\bd_{\nu}X^I\tq_{IJ}X_+^J}
{X_+^K\tq_{KL}X_+^L} \right) = \frac{p}{T_2\det g(X_+^K\tq_{KL}X_+^L)
(\det \tilde{Q})^{1/4}} \left( x^{\mu} - \frac{1}{2}\epsilon^{\mu\nu\la}
F_{\nu\la}\right),
\label{bp}\end{equation}
where
\begin{equation}
x^{\mu} = \pa^{\mu}X_+^IP_{IJ}X^J
\end{equation}
with a projection operator
\begin{equation}
P_{IJ} = \delta_{IJ} - \frac{X_+^IX_+^J}{X_+^2}.
\end{equation}
Substituting the expression (\ref{bp}) back into the starting Lagrangian
accompanied with the replacement (\ref{sp}) 
and solving the equation of motion for $p$ we get
\begin{eqnarray}
L &=& \st \Biggl[ - T_2 (\det\tilde{Q})^{1/4}\sqrt{-\det g}
\sqrt{ 1 + \frac{1}{2T_2 (X_+^K\tq_{KL}X_+^L)\sqrt{\det\tilde{Q}} }
\mathcal{F}_{\mu\nu}\mathcal{F}_{\rho\sigma}g^{\mu\rho}g^{\nu\sigma} }
\nonumber \\
&+& \frac{\bd_{\mu}X^I\tq_{IJ}X_+^J}{X_+^K\tq_{KL}X_+^L} \left( 
\frac{1}{2}\epsilon^{\mu\nu\la}F_{\nu\la} - x^{\mu} \right) \Biggr]
+ L_0,
\label{ep}\end{eqnarray}
where 
\begin{eqnarray}
\mathcal{F}_{\mu\nu} &=& F_{\mu\nu} - \frac{1}{\det g}\epsilon_{\mu\nu\la}
x^{\la},   \label{mf} \\
L_0 &=& \pa_{\mu}X_-^I\pa^{\mu}X_+^I - \tr\left( \frac{X_+\cdot X}{X_+^2}
D_{\mu}X^I\pa^{\mu}X_+^I - \frac{1}{2}\left(\frac{X_+\cdot X}{X_+^2}
\right)^2\pa_{\mu}X_+^I\pa^{\mu}X_+^I \right).
\end{eqnarray}
Here we use the identity for $3\times 3$ matrices $g_{\mu\nu} + 
a\mathcal{F}_{\mu\nu}$ with $\mathcal{F}_{\mu\nu} = -\mathcal{F}_{\nu\mu}$
\begin{equation}
\det(g_{\mu\nu})\left(1 + \frac{1}{2}a^2\mathcal{F}_{\mu\nu}
\mathcal{F}_{\rho\sigma}g^{\mu\rho}g^{\nu\sigma} \right) =
\det( g_{\mu\nu} + a\mathcal{F}_{\mu\nu} )
\label{de}\end{equation}
to obtain a DBI-type Lagrangian 
\begin{eqnarray}
L &=& \st \Biggl[ - T_2(\det\tilde{Q})^{1/4}\sqrt{-\det\left( g_{\mu\nu} +
\frac{1}{\sqrt{T_2 (X_+^K\tq_{KL}X_+^L)}(\det\tilde{Q})^{1/4} }
\mathcal{F}_{\mu\nu} \right)  } \nonumber \\
&+&  \frac{\bd_{\mu}X^I\tq_{IJ}X_+^J}{X_+^K\tq_{KL}X_+^L} \left( 
\frac{1}{2}\epsilon^{\mu\nu\la}F_{\nu\la} - x^{\mu} \right) \Biggr] + L_0.
\label{db}\end{eqnarray}

The inverse matrix of $\tilde{Q}^{IJ}$ in (\ref{qs}) is given by
\begin{equation}
\tq_{IJ} = P_{IJ} + \frac{X_+^IX_+^J}{X_+^2}\frac{1}{\det S} +
\left( \frac{m_0}{1 - m_0} \right)^{IJ},
\label{iq}\end{equation}
where an orthogonal relation $m^{IJ}X_+^J = 0$ is used and
\begin{eqnarray}
\left( \frac{m_0}{1 - m_0} \right)^{IJ} &=&  m_0^{IJ} + (m_0^2)^{IJ}
+ (m_0^3)^{IJ} + \cdots, \nonumber \\
m_0^{IJ} &=& - \frac{i}{\sqrt{T_2X_+^2}} m^{IJ}
\end{eqnarray}
with $(m_0^2)^{IJ} = m_0^{IK}m_0^{KJ}$. Since only the symmetric part of
matrix $\tq_{IJ}$ is taken into account in the Lagrangian, the 
expression (\ref{iq}) is modified to be
\begin{equation}
\tq_{IJ} = P_{IJ} + \frac{X_+^IX_+^J}{X_+^2}\frac{1}{\det S} +
\left( \frac{m_0^2}{1 - m_0^2} \right)^{IJ},
\label{qp}\end{equation}
which obeys $\tq_{IJ} = \tq_{JI}$ and
includes only terms with even number of $M^{IJK}$ as expressed by 
\begin{equation}
\left( \frac{m_0^2}{1 - m_0^2} \right)^{IJ} = (m_0^2)^{IJ}
+ (m_0^4)^{IJ} + (m_0^6)^{IJ} + \cdots.
\end{equation}
From this expression the following SO(8) invariant factors are 
simplified as
\begin{eqnarray}
X_+^I\tq_{IJ}X_+^J &=& X_+^2\frac{1}{\det S}, \nonumber  \\
\bd_{\mu}X^I\tq_{IJ}X_+^J &=& \bd_{\mu}X^IX_+^I\frac{1}{\det S}
\label{in}\end{eqnarray}
and the tensor
$\tilde{P}_{IJ}$ in (\ref{tp}) is also given by
\begin{equation}
\tilde{P}_{IJ} = \tq_{IJ} -  \frac{X_+^IX_+^J}{X_+^2}\frac{1}{\det S}.
\end{equation}
The relations in (\ref{in}) together with $\det\tilde{Q} = (\det S)^2$
make the DBI-type Lagrangian (\ref{db}) a simple form
\begin{eqnarray}
L &=& - T_2 \st\left( \sqrt{-\det\left( g_{\mu\nu} +
\frac{1}{\sqrt{T_2 X_+^2} }
\mathcal{F}_{\mu\nu} \right) } (\det S)^{1/2} \right) \nonumber \\
&+&  \tr \left(\frac{\bd_{\mu}X^IX_+^I}{X_+^2} \left( 
\frac{1}{2}\epsilon^{\mu\nu\la}F_{\nu\la} - x^{\mu} \right) \right) + L_0,
\label{si}\end{eqnarray}
where $g_{\mu\nu}$ defined in (\ref{tb}) is rewritten by
\begin{equation}
g_{\mu\nu} = \eta_{\mu\nu} + \frac{1}{T_2}\bd_{\mu}X^I\left( P_{IJ}
- \frac{1}{T_2X_+^2}\left( \frac{m^2}{1 + \frac{m^2}{T_2X_+^2}}
\right)^{IJ} \right)\bd_{\nu}X^J
\label{ge}\end{equation}
and there is a relation derived from (\ref{mf})
\begin{equation}
\frac{1}{2}\epsilon^{\mu\nu\la}\mathcal{F}_{\nu\la} = 
\frac{1}{2}\epsilon^{\mu\nu\la}F_{\nu\la}  - x^{\mu}.
\end{equation}
Thus from the non-linear BF Lagrangian with two nonpropagating
gauge fields $A_{\mu}, B_{\mu}$ we have integrated the auxiliary $B_{\mu}$
gauge field to extract the DBI-type Lagrangian expressed in terms of the
$SU(N)$ dynamical $A_{\mu}$ gauge field.

Now to perform the low-energy expansion for the non-linear 
Lagrangian (\ref{si}), we calculate $\det S^{IJ}$ for $8\times 8$
matrices by making the $1/T_2$ expansion as
\begin{equation}
\det S = 1 + \frac{1}{2T_2X_+^2}(m^2)^{II} - \frac{1}{4T_2^2(X_+^2)^2}
\left( (m^4)^{II} - \frac{1}{2}( (m^2)^{II} )^2 \right)
+ \cdots,
\label{ds}\end{equation}
where $m^{IJ} = - m^{JI}$ is taken into
account and $(m^2)^{II} = -X_+^2M^{IJK}M^{IJK}/3$.
There is the following identity with finite terms for any $3\times 3$
matrices $A_{\mu\nu}$
\begin{eqnarray}
\det(\eta_{\mu\nu} + A_{\mu\nu}) &=& \det\eta \Bigl( 1 + \mathrm{tr}
(\eta^{-1}A) - \frac{1}{2}\mathrm{tr}(\eta^{-1}A)^2 + \frac{1}{2}
\left(\mathrm{tr}(\eta^{-1}A)\right)^2 \nonumber \\
&+& \frac{1}{3}\mathrm{tr}(\eta^{-1}A)^3 - \frac{1}{2}\mathrm{tr}
(\eta^{-1}A)\mathrm{tr}(\eta^{-1}A)^2 \Bigr),
\end{eqnarray}
which gives the $1/T_2$ expansion for $\det g_{\mu\nu}$ in (\ref{ep})
\begin{eqnarray}
\det g_{\mu\nu} &=& - \Bigg( 1 + \frac{1}{T_2}\bd_{\mu}X^IP_{IJ}\bd^{\mu}
X^J + \frac{1}{T_2^2} \Bigl( - \frac{1}{2}\bd_{\mu}X^IP_{IJ}\bd_{\nu}X^J
\bd^{\nu}X^KP_{KL}\bd^{\mu}X^L \nonumber \\
&+&  \frac{1}{2}(\bd_{\mu}X^IP_{IJ}\bd^{\mu}X^J )^2  
- \frac{1}{X_+^2}\bd_{\mu}X^Im^{IK}m^{KJ}
\bd^{\mu}X^J \Bigr) + O\left(\frac{1}{T_2^3}\right) \Biggr).
\label{dg}\end{eqnarray}
We see that the $SO(8)$ vectors $\bd_{\mu}X^I$ are contracted with
$(m^2)^{IJ}$ and the projection operator $P^{IJ}$.
It is convenient to express the square root factor including
$\mathcal{F}_{\mu\nu}$ in (\ref{ep})
in terms of $F^{\mu} \equiv \epsilon^{\mu\nu\la}F_{\nu\la}/2 - x^{\mu}$
which appears as an interaction $\bd_{\mu}X^IX_+^IF^{\mu}/X_+^2$ in 
(\ref{si}), and expand it through (\ref{ge}) and (\ref{dg}) as
\begin{eqnarray}
\sqrt{1 + \frac{1}{2T_2X_+^2}\mathcal{F}_{\mu\nu}\mathcal{F}_{\rho\sigma}
g^{\mu\rho}g^{\nu\sigma} } = \sqrt{1 + \frac{1}{T_2X_+^2\det g}
F^{\mu}F^{\nu}g_{\mu\nu} } \nonumber \\
= 1 - \frac{1}{2T_2X_+^2}F^{\mu}F^{\nu}\eta_{\mu\nu} +
\frac{1}{2T_2^2X_+^2}\Biggl( F_{\mu}F^{\mu}\bd_{\nu}X^IP_{IJ}\bd^{\nu}X^J
\nonumber \\
- F^{\mu}F^{\nu}\bd_{\mu}X^IP_{IJ}\bd_{\nu}X^J 
- \frac{1}{4X_+^2}(F_{\mu}F^{\mu})^2 \Biggr) + O\left(\frac{1}{T_2^3}
\right).
\label{sf}\end{eqnarray}

Gathering  the expansions (\ref{ds}), (\ref{dg}) and (\ref{sf})
in (\ref{si}) or (\ref{ep}) we obtain the low-energy effective 
Lagrangian whose leading part is given by
\begin{equation}
L = -NT_2 + \tr \left( \frac{1}{12}M^{IJK}M^{IJK} - \frac{1}{2}
\bd_{\mu}X^IP_{IJ}\bd^{\mu}X^J + \frac{1}{2X_+^2}F_{\mu}F^{\mu}
+ \frac{1}{X_+^2}\bd_{\mu}X^IX_+^IF^{\mu} \right) + L_0,
\label{ll}\end{equation}
where $F_{\mu}F^{\mu}/2X_+^2$ is alternatively
expressed as $-f_{\mu\nu}f^{\mu\nu}
/4X_+^2$ in terms of $f_{\mu\nu} \equiv F_{\mu\nu} + 
\epsilon_{\mu\nu\la}x^{\la}$. This leading Lagrangian shows the Janus
field theory with a spacetime dependent coupling constant in ref. 
\cite{HIS} (see \cite{GW}).
This Lagrangian is rewritten by the following form 
\begin{eqnarray}
L = -NT_2 + \tr \Biggl( \frac{1}{12}M^{IJK}M^{IJK} - \frac{1}{2}
D_{\mu}X^IP_{IJ}D^{\mu}X^J + \frac{1}{2X_+^2}X^I\pa^{\mu}X_+^I
(X^J\pa_{\mu}X_+^J \nonumber \\
- 2D_{\mu}X^JX_+^J) - \frac{1}{4X_+^2}F_{\mu\nu}F^{\mu\nu} 
+ \frac{1}{2X_+^2}\epsilon^{\mu\nu\la}F_{\nu\la}(D_{\mu}X^IX_+^I 
- X^I\pa_{\mu}X_+^I) \Biggr) + \pa_{\mu}X_-^I\pa^{\mu}X_+^I,
\end{eqnarray}
which is further compactly represented by
\begin{eqnarray}
L &=& -NT_2 + \tr \Biggl( \frac{1}{12}M^{IJK}M^{IJK} - \frac{1}{2}
D_{\mu}X^ID^{\mu}X^I \nonumber \\
&+& \frac{1}{2X_+^2}
\left( \frac{1}{2}\epsilon^{\mu\nu\la}F_{\nu\la} + D^{\mu}X^IX_+^I
- X^I\pa^{\mu}X_+^I \right)^2 \Biggr) + \pa_{\mu}X_-^I\pa^{\mu}X_+^I.
\end{eqnarray}
The subleading terms of order $1/T_2$ are derived as
\begin{eqnarray}
\frac{1}{8T_2}\st \Biggl[ \frac{1}{(X_+^2)^2}m^{IJ}m^{JK}
m^{KL}m^{LI} -\frac{1}{36}((M^{IJK})^2)^2
+ 2\bd_{\mu}X^IP_{IJ}\bd_{\nu}X^J\bd^{\nu}X^KP_{KL}\bd^{\mu}X^L 
\nonumber \\
- (\bd_{\mu}X^IP_{IJ}\bd^{\mu}X^J)^2 
+ \frac{1}{3}(M^{IJK})^2\bd_{\mu}X^IP_{IJ}\bd^{\mu}X^J
+ \frac{4}{X_+^2}\bd_{\mu}X^Im^{IK}m^{KJ}\bd^{\mu}X^J \nonumber \\
+ \frac{(F_{\mu}F^{\mu})^2 }{(X_+^2)^2}
- \frac{F_{\mu}F^{\mu}}{3X_+^2}(M^{IJK})^2 
+ \frac{4F_{\mu}F_{\nu}}{X_+^2}\bd^{\mu}X^IP_{IJ}\bd^{\nu}X^J
- \frac{2F_{\mu}F^{\mu}}{X_+^2}\bd_{\nu}X^IP_{IJ}\bd^{\nu}X^J \Biggr].
\label{su}\end{eqnarray}
The last four terms including $F^{\mu} = \epsilon^{\mu\nu\la}f_{\nu\la}/2$
in (\ref{su}) are expressed in
terms of $f_{\mu\nu}$ as
\begin{eqnarray}
\frac{1}{8T_2X_+^2}\st \Biggl( \frac{1}{4X_+^2}(f_{\mu\nu}f^{\mu\nu})^2
+ 4f_{\mu\nu}f_{\rho\sigma}\eta^{\mu\rho}\bd^{\nu}X^IP_{IJ}\bd^{\sigma}X^J
\nonumber \\
- f_{\mu\nu}f^{\mu\nu}\bd_{\la}X^IP_{IJ}\bd^{\la}X^J 
+ \frac{1}{6}(M^{IJK})^2f_{\mu\nu}f^{\mu\nu} \Biggr),
\label{ff}\end{eqnarray}
where a $f_{\mu\nu}$ is accompanied with a factor $1/\sqrt{X_+^2}$.
The trace is taken symmetrically between all the matrix ingredients
$f_{\mu\nu}, \bd_{\mu}X^I, M^{IJK}$ so that the expression (\ref{ff})
is described by
\begin{eqnarray}
&\frac{1}{12T_2X_+^2}\tr \Biggl[ -\frac{1}{2}( 2 f_{\mu\nu}f^{\mu\nu}
\bd_{\la}X^I\bd^{\la}X^J + f_{\mu\nu}\bd_{\la}X^If^{\mu\nu}
\bd^{\la}X^J )P_{IJ} &   \\  
&+ \left(2(f_{\mu}^{\;\rho}f^{\mu\nu} + f_{\mu}^{\;\nu}f^{\mu\rho})
\bd_{\nu}X^I\bd_{\rho}X^J + f_{\mu}^{\;\rho}\bd_{\nu}X^If^{\mu\nu}
\bd_{\rho}X^J + f_{\mu}^{\;\nu}\bd_{\nu}X^If^{\mu\rho}\bd_{\rho}X^J
\right)P_{IJ}& \nonumber \\
&+ \frac{1}{8X_+^2}( 2 f_{\mu\nu}f^{\mu\nu}
f_{\rho\sigma}f^{\rho\sigma} + f_{\mu\nu}f_{\rho\sigma}
f^{\mu\nu}f^{\rho\sigma} ) + \frac{1}{12}( 2f_{\mu\nu}f^{\mu\nu}
(M^{IJK})^2 + f_{\mu\nu}M^{IJK}f^{\mu\nu}M^{IJK} )\Biggr].& \nonumber
\end{eqnarray}
The potential part in (\ref{su}) is also expanded as 
\begin{eqnarray}
&\frac{1}{24T_2}&\tr \Biggl[ \frac{1}{(X_+^2)^2}\left( (m^4)^{II}
+ 2(m^2)^{IJ}(m^2)^{IJ} \right) \nonumber \\
&-& \frac{1}{36}\left( 
M^{IJK}M^{LMN}M^{IJK}M^{LMN} + 2((M^{IJK})^2)^2 \right)\Biggr].
\end{eqnarray}
Here we write down the remaining terms
\begin{eqnarray}
&\frac{1}{12T_2}\tr \Biggl[ 
\bd_{\mu}X^I\bd_{\nu}X^J\bd^{\nu}X^K\bd^{\mu}X^L
 + \bd_{\mu}X^I\bd_{\nu}X^K\bd^{\nu}X^J
\bd^{\mu}X^L + \bd_{\mu}X^I\bd_{\nu}X^K\bd^{\mu}X^L\bd^{\nu}X^J& 
\nonumber \\
&- \bd_{\mu}X^I\bd^{\mu}X^J\bd_{\nu}X^K\bd^{\nu}X^L
- \frac{1}{2}\bd_{\mu}X^I\bd_{\nu}X^K\bd^{\mu}X^J\bd^{\nu}X^L
\Biggr]P_{IJ}P_{KL} &  \\
&+ \frac{1}{12T_2}\tr \Biggl[ \frac{2(m^2)^{IJ}}{X_+^2}
(\bd_{\mu}X^I\bd^{\mu}X^J + 
\bd_{\mu}X^J\bd^{\mu}X^I) - \frac{1}{X_+^2}( \bd_{\mu}X^Im^{IK}
\bd^{\mu}X^Jm^{JK}&  \nonumber \\
&+ m^{KI}\bd_{\mu}X^Im^{KJ}\bd^{\mu}X^J)
+\frac{1}{6}( 2(M^{LMN})^2\bd_{\mu}X^IP_{IJ}
\bd^{\mu}X^J + M^{LMN}\bd_{\mu}X^IM^{LMN}\bd^{\mu}X^JP_{IJ} ) \Biggr].&
\nonumber
\end{eqnarray}

\section{Two non-linear BF Lagrangians with even and odd number of 
$M^{IJK}$}

There are propositions of two types of non-linear BF Lagrangians
for multiple M2-branes, which include terms with even number as well as 
odd number of $M^{IJK}$ \cite{MG}.
One type is presented by
\begin{eqnarray}
L_1 &=& - T_2 \st\left( \sqrt{- \det\left(\eta_{\mu\nu} + 
\frac{1}{T_2}\td_{\mu}X^I\tilde{R}^{IJ}\td_{\nu}X^J \right) }
(\det S_1)^{1/4} \right) \nonumber \\
 &+& \frac{1}{2}\epsilon^{\mu\nu\la}\left( \tr(B_{\mu}F_{\nu\la})
- \frac{i}{T_2}\st( \td_{\mu}X^K\td_{\nu}X^IM^{IKN}(S_1^{-1})^{NJ}
\td_{\la}X^J) \right) \label{lo} \\
&+& (\pa_{\mu}X_-^I - \tr(X^IB_{\mu}))\pa^{\mu}X_+^I 
- \tr \left( \frac{X_+\cdot X}{X_+^2} \hat{D}_{\mu}X^I\pa^{\mu}X_+^I
- \frac{1}{2} \left(  \frac{X_+\cdot X}{X_+^2} \right)^2
\pa_{\mu}X_+^I\pa^{\mu}X_+^I \right), \nonumber
\end{eqnarray} 
where the symmetric tensor $\tilde{R}^{IJ}$ is defined by
\begin{eqnarray}
\tilde{R}^{IJ} &=&  (S_1^{-1})^{IJ} + \frac{X_+^IX_+^J}{X_+^2}
\left( \frac{1}{\sqrt{\det S_1}} - 1  \right), \nonumber \\
S_1^{IJ} &=& \delta^{IJ} - \frac{1}{T_2}M^{IKM}M^{JKN}
\left(\frac{X_+^MX_+^N}{X_+^2}\right).
\end{eqnarray}

Because of $\det S_1=(\det S)^2$ the symmetric tensor $\tilde{R}^{IJ}$
is identical to $\tq_{IJ}$ in (\ref{qp}), and $(\det S_1)^{1/4}
= (\det \tilde{Q})^{1/4}$, so that the Lagrangian $L_1$ except for 
terms with odd number of $M^{IJK}$ reduces to $L$ in (\ref{la}).
For this topological BF Lagrangian we consider the integration over
the $B_{\mu}$ gauge field to obtain a dynamical gauge theory Lagrangian.
Since the type one Lagrangian $L_1$  contains not only
the mass term of $B_{\mu}$ but also the cubic term, we cannot 
perform the $B_{\mu}$ integration directly. Instead, we begin to 
make the low-energy expansion for the non-linear term 
in (\ref{lo}) up to $1/T_2$ order 
\begin{eqnarray}
&-& T_2N + \st \Biggl[ -\frac{1}{2}\td_{\mu}X^I\td^{\mu}X^I + 
\frac{1}{4}A^{II} \nonumber \\
&+& \frac{1}{T_2} \left( - Z(B_{\mu}) + \frac{1}{8}( A^{II}\td_{\mu}X^J
\td^{\mu}X^J + A^{IJ}A^{JI} - \frac{1}{4}(A^{II})^2 ) \right) \Biggr],
\end{eqnarray}
where 
\begin{eqnarray}
A^{IJ} &=& M^{IKM}M^{JKN}\left( \frac{X_+^MX_+^N}{X_+^2} \right)
= - \frac{1}{X_+^2}(m^2)^{IJ}, \nonumber \\
Z(B_{\mu}) &=& \frac{1}{8}\Biggl( ( \td_{\mu}X^I\td^{\mu}X^I)^2 
- 2\td_{\mu}X^I\td_{\nu}X^I \td^{\nu}X^J\td^{\mu}X^J \nonumber \\
&+& 4\td_{\mu}X^I\left(A^{IJ} + \frac{X_+^IX_+^J}{2X_+^2}A^{KK}
\right) \td^{\mu}X^J \Biggr).
\end{eqnarray}

The algebraic equation of motion for $B_{\mu}$ reads
\begin{eqnarray}
X_+^I(\bd^{\mu}X^I &-& X_+^IB^{\mu} ) + \frac{1}{2}\epsilon^{\mu\nu\la}
F_{\nu\la} - x^{\mu} \nonumber \\
&=& \frac{1}{T_2} 
\left(\frac{1}{4}A^{II}X_+^J(\bd^{\mu}X^J - X_+^JB^{\mu})
+ \frac{\delta Z}{\delta B_{\mu}} + \frac{i}{2}\epsilon^{\rho\nu\la}
\frac{\delta X_{\rho\nu\la}}{\delta B_{\mu}} \right)
\end{eqnarray}
with $X_{\rho\nu\la}(B^{\mu}) = \td_{\rho}X^K\td_{\nu}X^IM^{IKJ}
\td_{\la}X^J$. The solution can be iteratively derived by
$B^{\mu} = B_0^{\mu} + B_1^{\mu}/T_2$, with
\begin{equation}
B_0^{\mu} = \frac{1}{X_+^2}( X_+^I\bd^{\mu}X^I + \frac{1}{2}
\epsilon^{\mu\nu\la}F_{\nu\la} - x^{\mu})
\label{bx}\end{equation}
and
\begin{equation}
B_1^{\mu} = -\frac{1}{X_+^2}\left( \frac{1}{4}A^{II}X_+^J(\bd^{\mu}X^J 
- X_+^JB_0^{\mu}) + \frac{\delta Z}{\delta B_{\mu}}\Big|_{B_0^{\mu}}
 + \frac{i}{2}\epsilon^{\rho\nu\la}\frac{\delta X_{\rho\nu\la}}
{\delta B_{\mu}}\Big|_{B_0^{\mu}} \right),
\end{equation}
where the expression of $B_0^{\mu}$ (\ref{bx}) is inserted into
the last two derivative terms. Substituting this solution back 
into the low-energy Lagrangian of $L_1$ (\ref{lo}) we obtain
the same leading Lagrangian as (\ref{ll}) through a relation 
\begin{equation}
\bd^{\mu}X^I - X_+^IB_0^{\mu} = P^{IJ}\bd^{\mu}X^J - \frac{1}{X_+^2}
X_+^IF^{\mu}
\label{dp}\end{equation}
and the following 
correction terms of order $1/T_2$
\begin{eqnarray}
\frac{1}{T_2} \st \Biggl( (\bd_{\mu}X^I - X_+^IB_{0\mu} )X_+^IB_1^{\mu}
+ \frac{1}{8}A^{II} (\bd_{\mu}X^J - X_+^JB_{0\mu} )
(\bd^{\mu}X^J - X_+^JB_0^{\mu} ) \nonumber \\
+ \frac{1}{8}( A^{IJ}A^{JI} - \frac{1}{4}(A^{II})^2 ) - Z(B_0^{\mu})
+ F^{\mu}B_{1\mu} - \frac{i}{2}\epsilon^{\mu\nu\la}
X_{\mu\nu\la}(B_{0}^{\mu}) \Biggr).
\end{eqnarray}
The subleading terms except for the terms including 
$X_{\mu\nu\la}(B_0^{\mu})$
and $\delta X_{\rho\nu\la}/\delta B_{\mu}|_{B_0^{\mu}}$
are described by
\begin{eqnarray}
&\frac{1}{T_2}&\st \Biggl[ \frac{1}{8}\left( A^{KK}\bd_{\mu}X^IP_{IJ}
\bd^{\mu}X^J - A^{II}\frac{F_{\mu}F^{\mu}}{X_+^2} + A^{IJ}A^{JI} 
- \frac{1}{4}(A^{II})^2 \right) \nonumber \\
&+& \frac{1}{4} \left( \bd_{\mu}X^IP_{IJ}\bd_{\nu}X^J
\bd^{\nu}X^KP_{KL}\bd^{\mu}X^L + \frac{2F_{\mu}F_{\nu}}{X_+^2}\bd^{\mu}
X^IP_{IJ}\bd^{\nu}X^J + \frac{(F_{\mu}F^{\mu})^2}{(X_+^2)^2}
\right) \nonumber \\
&-& \frac{1}{8} \left( (\bd_{\mu}X^IP_{IJ}\bd^{\mu}X^J)^2
 + \frac{2F_{\mu}F^{\mu}}{X_+^2}\bd_{\nu}X^IP_{IJ}\bd^{\nu}X^J 
+ \frac{(F_{\mu}F^{\mu})^2}{(X_+^2)^2}
\right) \nonumber \\
&+& \frac{1}{2X_+^2} \bd_{\mu}X^Im^{IK}m^{KJ}\bd^{\mu}X^J \Biggr].
\label{sl}\end{eqnarray}
It is noted that the $SO(8)$ vectors $\bd_{\mu}X^I$
are contracted with $(m^2)^{IJ}$
and the projection operator $P_{IJ}$ which is due to (\ref{dp}). 
The expression (\ref{sl}) is confirmed to agree with (\ref{su}).
Thus we have observed that these subleading
 terms  obtained by the iterative 
procedure for the $B_{\mu}$ integration in the low-energy Lagrangian
reproduces the previous expression (\ref{su}) which is derived by
the low-energy expansion of the effective DBI-type Lagrangian
generated by the exact $B_{\mu}$ integration.

The remaining terms lead to
\begin{eqnarray}
&-& \frac{i}{2T_2} \st \left( \epsilon^{\mu\nu\la}X_{\mu\nu\la}(B_0^{\mu})
+ \epsilon^{\rho\nu\la}\frac{\delta X_{\rho\nu\la}}{\delta B_{\mu}}
\Big|_{B_0^{\mu}} \left( \frac{X_+^I}{X_+^2}(\bd_{\mu}X^I - X_+^IB_{0\mu})
+ \frac{1}{X_+^2}F_{\mu} \right) \right) \nonumber \\
&=& \frac{i}{2T_2}\epsilon^{\mu\nu\la}\st \left( \tilde{M}^{IJK}
\bd_{\mu}X^I\bd_{\nu}X^J\bd_{\la}X^K - \frac{3m^{IJ}}{X_+^2}
F_{\mu}\bd_{\nu}X^I\bd_{\la}X^J \right),
\label{er}\end{eqnarray}
where $\tilde{M}^{IJK}$ is a totally antisymmetric tensor defined by
\begin{equation}
\tilde{M}^{IJK} = M^{IJK} - \frac{1}{X_+^2}( m^{IJ}X_+^K +
m^{JK}X_+^I + m^{KI}X_+^J ),
\end{equation}
which is orthogonal to $X_+^{I}$ as $\tilde{M}^{IJK}X_+^I=0$.
This expression including single $M^{IJK}$ is rewritten by
\begin{equation}
\frac{i}{2T_2} \st (\epsilon^{\mu\nu\la}\tilde{M}^{IJK}
\bd_{\mu}X^I\bd_{\nu}X^J\bd_{\la}X^K + 3m^{IJ}f^{\mu\nu}
\bd_{\mu}X^I\bd_{\nu}X^J ).
\label{se}\end{equation}
We see that there is a coupling between the gauge field strength
$F^{\mu\nu}$ and $m^{IJ}\bd_{\mu}X^I\bd_{\nu}X^J$.
The symmetrization in (\ref{se}) is taken as
\begin{eqnarray}
\frac{i}{2T_2} \tr (\epsilon^{\mu\nu\la}\tilde{M}^{IJK}
\bd_{\mu}X^I\bd_{\nu}X^J\bd_{\la}X^K + m^{IJ}f^{\mu\nu}
\bd_{\mu}X^I\bd_{\nu}X^J \nonumber \\
+ m^{IJ}\bd_{\mu}X^If^{\mu\nu}\bd_{\nu}X^J + m^{IJ}\bd_{\mu}X^I
\bd_{\nu}X^J f^{\mu\nu} ),
\end{eqnarray}
where the first term remains intact. 

Now we turn to the other type of non-linear Lagrangian for non-Abelian
BF membranes
\begin{eqnarray}
L_2&=& - T_2 \st\left( \sqrt{- \det\left(\eta_{\mu\nu} + 
\frac{1}{T_2}\td_{\mu}X^I(S_2^{-1})^{IJ}\td_{\nu}X^J \right) }
(\det S_2)^{1/6} \right) \nonumber \\
 &+& \frac{1}{2}\epsilon^{\mu\nu\la}\left( \tr(B_{\mu}F_{\nu\la})
- \frac{i}{T_2}\st( \td_{\mu}X^K\td_{\nu}X^IM^{IKN}(S_2^{-1})^{NJ}
\td_{\la}X^J) \right)  \label{lt}  \\
&+& (\pa_{\mu}X_-^I - \tr(X^IB_{\mu}))\pa^{\mu}X_+^I 
- \tr \left( \frac{X_+\cdot X}{X_+^2} \hat{D}_{\mu}X^I\pa^{\mu}X_+^I
- \frac{1}{2} \left(  \frac{X_+\cdot X}{X_+^2} \right)^2
\pa_{\mu}X_+^I\pa^{\mu}X_+^I \right), \nonumber 
\end{eqnarray}
where the symmetric tensor $S_2^{IJ}$ is defined by
\begin{equation}
S_2^{IJ} = \delta^{IJ} - \frac{1}{2T_2}B^{IJ}
\end{equation}
with $B^{IJ} = M^{IKM}M^{JKM} \equiv (M^2)^{IJ}$.
We make the following low-energy expansion for the non-linear
term in (\ref{lt})
\begin{eqnarray}
&-& T_2N + \st \Biggl[ -\frac{1}{2}\td_{\mu}X^I\td^{\mu}X^I + 
\frac{1}{12}B^{II} \nonumber \\
&+& \frac{1}{T_2} \left( - W(B_{\mu}) + \frac{1}{24}\left( B^{II}
\td_{\mu}X^J\td^{\mu}X^J + \frac{1}{2} B^{IJ}B^{JI} - \frac{1}{12}
(B^{II})^2  \right) \right) \Biggr],
\end{eqnarray}
where
\begin{equation}
W(B_{\mu}) = \frac{1}{8}\left( ( \td_{\mu}X^I\td^{\mu}X^I)^2 
- 2\td_{\mu}X^I\td_{\nu}X^I \td^{\nu}X^J\td^{\mu}X^J
+ 2\td_{\mu}X^IB^{IJ}\td^{\mu}X^J \right).
\end{equation}

The algebraic equation of motion for $B_{\mu}$ is also given by
\begin{eqnarray}
&X_+^I&(\bd^{\mu}X^I - X_+^IB^{\mu} ) + \frac{1}{2}\epsilon^{\mu\nu\la}
F_{\nu\la} - x^{\mu} \nonumber \\
&=& \frac{1}{T_2}\left(\frac{1}{12}
B^{II}X_+^J(\bd^{\mu}X^J - X_+^JB^{\mu})
+ \frac{\delta W}{\delta B_{\mu}} + \frac{i}{2}\epsilon^{\rho\nu\la}
\frac{\delta X_{\rho\nu\la}}{\delta B_{\mu}} \right),
\end{eqnarray}
whose solution is iteratively derived by
$B^{\mu} = B_0^{\mu} + \tilde{B}_1^{\mu}/T_2$ where $B_0^{\mu}$ is the 
same expression as (\ref{bx}) and $\tilde{B}_1^{\mu}$ is
\begin{equation}
\tilde{B}_1^{\mu} = -\frac{1}{X_+^2}\left( \frac{1}{12}B^{II}X_+^J
(\bd^{\mu}X^J - X_+^JB_0^{\mu}) + \frac{\delta W}{\delta B_{\mu}}
\Big|_{B_0^{\mu}}  + \frac{i}{2}\epsilon^{\rho\nu\la}
\frac{\delta X_{\rho\nu\la}}{\delta B_{\mu}}\Big|_{B_0^{\mu}} \right).
\end{equation}
The substitution of this solution into the low-energy Lagrangian of 
$L_2$ (\ref{lt}) yields the same leading Lagrangian as (\ref{ll})
and the following subleading terms of order $1/T_2$
\begin{eqnarray}
\frac{1}{T_2} \st \Biggl( (\bd_{\mu}X^I - X_+^IB_{0\mu} )X_+^I
\tilde{B}_1^{\mu} + \frac{1}{24}B^{II} (\bd_{\mu}X^J - X_+^JB_{0\mu} )
(\bd^{\mu}X^J - X_+^JB_0^{\mu} ) \nonumber \\
+ \frac{1}{48}( B^{IJ}B^{JI} - \frac{1}{6}(B^{II})^2 ) - W(B_0^{\mu})
+ F^{\mu}\tilde{B}_{1\mu} - \frac{i}{2}\epsilon^{\mu\nu\la}
X_{\mu\nu\la}(B_{0}^{\mu}) \Biggr).
\label{sb}\end{eqnarray}
We use an identity $B^{II} = 3A^{II}$ to express (\ref{sb}) as sum
of (\ref{er}) and 
\begin{eqnarray}
&\frac{1}{T_2}&\st \Biggl[ \frac{1}{8}\left( A^{KK}\bd_{\mu}X^IP_{IJ}
\bd^{\mu}X^J - A^{II}\frac{F_{\mu}F^{\mu}}{X_+^2} + 
\frac{1}{6}(M^2)^{IJ}(M^2)^{JI}
 - \frac{1}{4}(A^{II})^2 \right) \nonumber \\
&+& \frac{1}{4} \left( \bd_{\mu}X^IP_{IJ}\bd_{\nu}X^J
\bd^{\nu}X^KP_{KL}\bd^{\mu}X^L + \frac{2F_{\mu}F_{\nu}}{X_+^2}\bd^{\mu}
X^IP_{IJ}\bd^{\nu}X^J + \frac{(F_{\mu}F^{\mu})^2}{(X_+^2)^2}
\right) \nonumber \\
&-& \frac{1}{8} \left( (\bd_{\mu}X^IP_{IJ}\bd^{\mu}X^J)^2
 + \frac{2F_{\mu}F^{\mu}}{X_+^2}\bd_{\nu}X^IP_{IJ}\bd^{\nu}X^J 
+ \frac{(F_{\mu}F^{\mu})^2}{(X_+^2)^2}
\right) \nonumber \\
&-& \frac{1}{4} \bd_{\mu}X^I\bar{M}^{IMN}\bar{M}^{JMN}\bd^{\mu}X^J
 \Biggr],
\end{eqnarray}
where $\bar{M}^{IMN} = P^{IK}M^{KMN}$ and a relation
$\bar{M}^{IMN}m^{MN} = 0$ is used.
The $1/T_2$ corrections show almost similar expressions to (\ref{sl})
with two different terms which are a potential term 
$(M^2)^{IJ}(M^2)^{JI}/6$ and an interaction term 
$-\bd_{\mu}X^I\bar{M}^{IMN}\bar{M}^{JMN}\bd^{\mu}X^J/4$.

\section{Conclusion}

Without resort to the low-energy expansion we have performed the
integration over one Chern-Simons nonpropagating $B_{\mu}$ gauge field
exactly for the non-linear Lagrangian of the BF membrane theory
in ref. \cite{IR}, which includes terms with even number of
$M^{IJK}$. We have observed that there appears a non-linear
DBI-type Lagrangian for the worldvolume theory of $N$ M2-branes
where the other Chern-Simons $A_{\mu}$ gauge field is promoted to the
$SU(N)$ dynamical propagating gauge field. 

In the  non-linear DBI-type Lagrangian the coefficient factor
of the modified field strength 
$\mathcal{F}_{\mu\nu}$ takes a compact form $1/\sqrt{T_2X_+^2}$
which yields the kinetic term of gauge field 
$-F_{\mu\nu}F^{\mu\nu}/4X_+^2$ with a space-time dependent coupling
field $X_+^I$ in the leading low-energy expansion. 
In the same way the linear term of $F^{\mu}$ also takes a compact
interaction $\bd_{\mu}X^IX_+^IF^{\mu}/X_+^2$. The subleading
terms including dynamical gauge field strength $F_{\mu\nu}$
are expressed in terms of $F^{\mu}$ or alternatively $f_{\mu\nu}$
which is a specific combination of $F_{\mu\nu}$ and an $SO(8)$
invariant contraction of scalar fields with the projection
operator  $\pa^{\mu}X_+^IP_{IJ}X^J$. In the subleading
terms the $SO(8)$ vectors $\bd_{\mu}X^I$ 
are contracted with $P^{IJ}$ and
$(m^2)^{IJ}$ consisting of two $M^{IJK}$, which are 
orthogonal to $X_+^I$.  This Lagrangian is regarded as the
non-linear extension of the Janus field theory Lagrangian
in ref. \cite{HIS}.

For the two types of non-linear BF Lagrangians 
in ref. \cite{MG} which include terms with even and odd
number of $M^{IJK}$, we have made the low-energy expansion
and then carried out the $B_{\mu}$ integration by solving
its equation of motion in the presence of the $1/T_2$ order
corrections through an iterative procedure.
In the type one Lagrangian $L_1$ we have demonstrated that
there appear indeed various terms at order $1/T_2$
in the iteratively $B_{\mu}$ integrated effective Lagrangian, 
but they except for terms with odd number of
$M^{IJK}$ are reshuffled to be in agreement with the $1/T_2$
order terms in the low-energy expansion of the above
exactly $B_{\mu}$ integrated Lagrangian of the DBI form.
In the type two Lagrangian $L_2$ the effective Lagrangian has
been observed to have almost similar $1/T_2$ order corrections
except for two terms, where $\bd_{\mu}X^I$ are contracted with
$P^{IJ}$ as well as $\bar{M}^{IMN}\bar{M}^{JMN}$ which is also
orthogonal to $X_+^I$.
The remaining terms including single $M^{IJK}$ in both
effective Lagrangians consist of two kinds of interactions 
specified by $\epsilon^{\mu\nu\la}\tilde{M}^{IJK}
\bd_{\mu}X^I\bd_{\nu}X^J\bd_{\la}X^K$ and $m^{IJ}f^{\mu\nu}
\bd_{\mu}X^I\bd_{\nu}X^J$ where the SO(8) vectors $\bd_{\mu}X^I$
are contracted with the tensors $\tilde{M}^{IJK}$ and $m^{IJ}$
which are orthogonal to $X_+^I$.


\begin{thebibliography}{99}
\bibitem{BL} J. Bargger and N. Lambert, ``Modeling multiple M2's,"
Phys. Rev. \textbf{D75}, 045020 (2007) [arXiv:hep-th/0611108];
``Gauge symmetry and supersymmetry of multiple M2-branes,"
Phys. Rev. \textbf{D77}, 065008 (2008) [arXiv:0711.0955[hep-th]];
``Comments on multiple M2-branes," JHEP \textbf{0802},105 (2008)
[arXiv:0712.3738[hep-th]].
\bibitem{AG} A. Gustavsson, ``Algebraic structures on parallel
M2-branes," Nucl. Phys. \textbf{B811}, 66 (2009)
[arXiv:0709.1260[hep-th]];
``Selfdual strings and loop space Nahm equations," JHEP \textbf{0804},
083 (2008) [arXiv:0802.3456[hep-th]].
\bibitem{JS} J.H. Schwarz, ``Superconformal Chern-Simons theories," 
JHEP \textbf{0411}, 078 (2004) [arXiv:hep-th/0411077].
\bibitem{BH} A. Basu and J.A. Harvey, ``The M2-M5 brane system and
a generalized Nahm's equation," Nucl. Phys. \textbf{B713}, 136
(2005) [arXiv:hep-th/0412310].
\bibitem{GP} G. Papadopoulos, ``M2-branes, 3-Lie algebras and Plucker
relations," JHEP \textbf{0805}, 054 (2008) [arXiv:0804.2662[hep-th]];
J.P. Gauntlett and J.B. Gutowski, ``Constraining maximally supersymmetric
membrane actions," arXiv:0804.3078[hep-th].
\bibitem{MR} M. Van Raamsdonk, ``Comments on the Bagger-Lambert theory
and multiple M2-branes," JHEP \textbf{0805}, 105 (2008) 
[arXiv:0803.3803[hep-th]];
N. Lambert and D. Tong, ``Membranes on an orbifold," 
Phys. Rev. Lett. \textbf{101}, 041602 (2008) [arXiv:0804.1114[hep-th]];
J. Distler, S. Mukhi, C. Papageorgakis and M. Van Raamsdonk, 
``M2-branes on M-folds," JHEP \textbf{0805}, 038 (2008) 
[arXiv:0804.1256[hep-th]].
\bibitem{GMR} J. Gomis, G. Milanesi and J.G. Russo, ``Bagger-Lambert
theory for general Lie algebras," JHEP \textbf{0806}, 075 (2008) 
[arXiv:0805.1012[hep-th]].
\bibitem{BGTV} S. Benvenuti, D. Rodriguez-Gomez, E. Tonni and 
H. Verlinde, ``$\mathcal{N} = 8$ superconformal gauge theories and
M2 branes," JHEP \textbf{0901}, 078 (2009) [arXiv:0805.1087[hep-th]].
\bibitem{HIM} P.-M. Ho, Y. Imamura and Y. Matsuo, ``M2 to D2 revisited,"
JHEP \textbf{0807}, 003 (2008) [arXiv:0805.1202[hep-th]].
\bibitem{MP} S. Mukhi and C. Papageorgakis, ``M2 to D2,"
JHEP \textbf{0805}, 085 (2008) [arXiv:0803.3218[hep-th]].
\bibitem{HIS} Y. Honma, S. Iso, Y. Sumitomo and S. Zhang, ``Janus
field theories from multiple M2 branes," 
Phys. Rev. \textbf{D78}, 025027 (2008) [arXiv:0805.1895[hep-th]].
\bibitem{BLS} M.A. Bandres, A.E. Lipstein and J.H. Schwarz, 
``Ghost-free superconformal action for multiple M2-branes," 
JHEP \textbf{0807}, 117 (2008) [arXiv:0806.0054[hep-th]].
\bibitem{GGRV} J. Gomis, D. Rodriguez-Gomez, M. Van Raamsdonk
and H. Verlinde, ``Supersymmetric Yang-Mills theory from
Lorentzian three-algebras," 
JHEP \textbf{0808}, 094 (2008) [arXiv:0806.0738[hep-th]].
\bibitem{HV} H. Verlinde, ``D2 or M2 ? A note on membrane 
scattering," arXiv:0807.2121[hep-th].
\bibitem{EMP} B. Ezhuthachan, S. Mukhi and C. Papageorgakis,
``D2 to D2," JHEP \textbf{0807}, 041 (2008) [arXiv:0806.1639[hep-th]].
\bibitem{NS} H. Nicolai and H. Samtleben, ``Chern-Simons vs.
Yang-Mills gaugings in three dimensions," Nucl. Phys. 
\textbf{B668}, 167 (2003) [arXiv:hep-th/0303213];
B. de Wit, H. Nicolai and H. Samtleben, ``Gauged supergravities
in three dimensions: A panoramic overview," arXiv:hep-th/0403014.
\bibitem{ABJM} O. Aharony, O. Bergman, D.L. Jafferis and J. Maldacena,
``$\mathcal{N}=6$ superconformal Chen-Simons-matter theories,
M2-branes and their gravity duals," JHEP \textbf{0810}, 091 (2008)
[arXiv:0806.1218[hep-th]].
\bibitem{HISZ} Y. Honma, S. Iso, Y. Sumitomo and Z. Zhang, 
``Scaling limit of $\mathcal{N}=6$ superconformal Chern-Simons
theories and Lorentzian Bagger-Lambert theories," 
Phys. Rev. \textbf{D78}, 105011 (2008) [arXiv:0806.3498[hep-th]];
``Generalized conformal symmetry and recovery of SO(8) 
in multiple M2 and D2 branes," arXiv:0807.3825[hep-th].
\bibitem{JBL} J. Bagger and N. Lambert, ``Three-algebras and 
$\mathcal{N}=6$ Chern-Simons gauge theories," 
Phys. Rev. \textbf{D79}, 025002 (2009) [arXiv:0807.0163[hep-th]].
\bibitem{PW} Y. Pang and T. Wang, ``From $N$ M2's to $N$ D2's,"
Phys. Rev. \textbf{D78}, 125007 (2008) [arXiv:0807.1444[hep-th]].
\bibitem{AT} E. Antonyan and A.A. Tseytlin, ``On 3d $\mathcal{N}=8$
Lorentzian BLG theory as a scaling limit of 3d superconformal
$\mathcal{N}=6$ ABJM theory," 
Phys. Rev. \textbf{D79}, 046002 (2009) [arXiv:0811.1540[hep-th]].
\bibitem{HHM} P.-M. Ho, R.-C. Hou and Y. Matsuo, ``Lie 3-algebra
and multiple M2-branes," JHEP \textbf{0806}, 020 (2008) 
[arXiv:0804.2110[hep-th]];
P.-M. Ho and Y. Matsuo, ``M5 from M2," JHEP \textbf{0806}, 105 (2008)
[arXiv:0804.3629[hep-th]];
P.-M. Ho, Y. Imamura Y. Matsuo and S. Shiba, ``M5-brane in three-form
flux and multiple M2-branes," JHEP \textbf{0808}, 014 (2008) 
[arXiv:0805.2898[hep-th]];
J.H. Park and C. Sochichiu, ``Single M5 to multiple M2:
taking off the square root of Nambu-Goto action," 
arXiv:0806.0335[hep-th];
I.A. Bandos and P.K. Townsend, ``Light-cone M5 and multiple 
M2-branes," Class. Quant. Grav. \textbf{25}, 245003 (2008)
[arXiv:0806.4777[hep-th]];
``SDiff gauge theory and the M2 condensate," 
JHEP \textbf{0902}, 013 (2009) [arXiv:0808.1583[hep-th]].
\bibitem{AM} A. Morozov, ``On the problem of multiple M2 branes,"
JHEP \textbf{0805}, 076 (2008) [arXiv:0804.0913[hep-th]];
U. Gran, B.E.W. Nilsson and C. Petersson, ``On relating multiple
M2 and D2-branes," JHEP \textbf{0810}, 067 (2008) 
[arXiv:0804.1784[hep-th]];
E.A. Bergshoeff, M. de Roo and O. Hohm, ``Multiple M2-branes and
the embedding tensor," Class. Quant. Grav. \textbf{25}, 142001
(2008) [arXiv:0804.2201[hep-th]];
S. Banerjee and A. Sen, ``Interpreting the M2-brane action," 
arXiv:0805.3930[hep-th];
S. Cecotti and A. Sen, ``Coulomb branch of the Lorentzian
three algebra theory," arXiv:0806.1990[hep-th];
E.A. Bergshoeff, M. de Roo, O. Hohm and D. Roet, ``Multiple 
membranes from gauged supergravity," JHEP \textbf{0808}, 091 
(2008) [arXiv:0806.2584[hep-th]].
\bibitem{HL} H. Lin, ``Kac-Moody extensions of 3-algebras and 
M2-branes," JHEP \textbf{0807}, 136 (2008) [arXiv:0805.4003[hep-th]];
P. de Medeiros, J. Figueroa-O'Farrill and E. Mendez-Escobar,
``Lorentzian Lie 3-algebras and their Bagger-Lambert moduli space,"
JHEP \textbf{0807}, 111 (2008) [arXiv:0805.4363[hep-th]];
``Metric Lie 3-algebras in Bagger-Lambert theory,"
JHEP \textbf{0808}, 045 (2008) [arXiv:0806.3242[hep-th]];
M. Ali-Akbari, M.M. Sheikh-Jabbari and J. Simon, ``The relaxed
three-algebras: their matrix representations and implications
for multi M2-brane theory," JHEP \textbf{0812}, 037 (2008) 
[arXiv:0807.1570[hep-th]];
S. Cherkis and C. Saemann, ``Multiple M2-branes and generalized 
3-Lie algebra," Phys. Rev. \textbf{D78}, 066019 (2008)
[arXiv:0807.0808[hep-th]];
S. Cherkis, V. Dotsenko and C. Saemann, ``On superspace actions for
multiple M2-branes, metric 3-algbras and their classification," 
arXiv:0812.3127[hep-th];
C.I. Lazroiu, D. McNamee, C. Saemann and A. Zejak, ``Strong homotopy
Lie algebras, generalized Nahm equations and multiple M2-branes," 
arXiv:0901.3905[hep-th].
\bibitem{HMS} P.-M. Ho, Y. Matsuo and S. Shiba, ``Lorentzian Lie
 (3-)algebra and toroidal compactification of M/string theory," 
JHEP \textbf{0903}, 045 (2009) [arXiv:0901.2003[hep-th]];
P. de Medeiros, J. Figueroa-O'Farrill, E. Mendez-Escobar and 
P. Ritter, ``Metric 3-Lie algebras for unitary Bagger-Lambert theories,"
arXiv:0902.4674[hep-th].
\bibitem{IR} R. Iengo and J.G. Russo, ``Non-linear theory for multiple
M2 branes," JHEP \textbf{0810}, 030 (2008) [arXiv:0808.2473[hep-th]].
\bibitem{AAT} A.A. Tseytlin, ``On non-Abelian generalisation of
the Born-Infeld action in string theory," Nucl. Phys. \textbf{B501}, 41
(1997) [arXiv:hep-th/9701125].
\bibitem{RM} R.C. Myers, ``Dielectric-branes," JHEP \textbf{9912}, 022 
(1999) [arXiv:hep-th/9910053].
\bibitem{MG} M.R. Garousi, ``On non-linear action of multiple M2-branes,"
Nucl. Phys. \textbf{B809}, 519 (2009) [arXiv:0809.0985[hep-th]].
\bibitem{JK} J. Kluson, ``D2 to M2 procedure for D2-brane DBI effective
action," Nucl. Phys. \textbf{B808}, 260 (2009) [arXiv:0807.4054[hep-th]].
\bibitem{MAM} M. Alishahiha and S. Mukhi, ``Higher-derivative 3-algebras,"
JHEP \textbf{0810}, 032 (2008) [arXiv:0808.3067[hep-th]].
\bibitem{LLX} T. Li, Y. Liu and D. Xie, ``Multiple D2-brane action from
M2-branes," arXiv:0807.1183[hep-th].
\bibitem{BEMP} B. Ezhuthachan, S. Mukhi and C. Papageorgakis, ``The power
of the Higgs mechanism: higher-derivative 
BLG theories," arXiv:0903.0003[hep-th].
\bibitem{LW} M. Li and T. Wang, ``M2-branes coupled to antisymmetric 
fluxes," JHEP \textbf{0807}, 093 (2008) [arXiv:0805.3427[hep-th]].
\bibitem{MGA} M.A. Ganjali, ``Nambu-Poisson bracket and M-theory branes 
coupled to antisymmetric fluxes," arXiv:0811.2976[hep-th].
\bibitem{MAG} M.A. Ganjali, ``On dielectric membranes," 
arXiv:0901.2642[hep-th].
\bibitem{MRG} M.R. Garousi, ``A proposal for M2-brane-anti-M2-bane 
action," arXiv:0809.0381[hep-th].
\bibitem{JJK} J. Kluson, ``Note about unstable M3-brane action," 
Phys. Rev. \textbf{D79}, 026001 (2009) [arXiv:0810.0585[hep-th]].
\bibitem{GW} D. Gaiotto and E. Witten, ``Janus configurations,
Chern-Simons couplings, and the theta-angle in $\mathcal{N}=4$
super Yang-Mills theory," arXiv:0804.2907[hep-th];
K. Hosomichi, K.M. Lee, S. Lee and J. Park, ``$\mathcal{N}=4$
superconformal Chern-Simons theories with hyper and twisted hyper
multiplets," JHEP \textbf{0807}, 091 (2008) [arXiv:0805.3662[hep-th]].


\end{thebibliography}
\end{document}